\documentclass[twocolumn,showpacs,preprintnumbers,amsmath,amssymb]{revtex4}

\begin{document}

\preprint{}

\title{%
\boldmath
Spectroscopic distinction between the normal state pseudogap and the
superconducting gap of cuprate high\ T$_{c}$ superconductors
\unboldmath
}

\author{Li Yu$^{1,2}$, D. Munzar$^{3}$, A.V. Boris$^{2}$, P. Yordanov$^{2}$,
J. Chaloupka$^{3}$, Th. Wolf$^{4}$, C.T. Lin$^{2}$, B. Keimer$^{2}$, and C.
Bernhard$^{1,2}$}
\affiliation{1.) University of Fribourg, Department of Physics and Fribourg, Center for Nanomaterials (FriMat), Chemin du Musee 3, CH-1700 Fribourg, Switzerland\\
2.) Max-Planck-Institute for Solid State Research, Heisenbergstrasse 1, D-70569 Stuttgart, Germany\\ 
3.) Institute of Condensed Matter Physics, Masaryk University, Kotl{\'{a}}rsk%
{\'{a}} 2, CZ-61137 Brno, Czech Republic\\
4.) Forschungszentrum Karlsruhe, IFP, D-76021 Karlsruhe, Germany}

\date{\today}

\begin{abstract} We report on broad-band infrared ellipsometry measurements of the c-axis
conductivity of underdoped RBa$_{2}$Cu$_{3}$O$_{7-d}$ (R=Y, Nd, and La)
single crystals. Our data provide a detailed account of the spectral weight
(SW) redistributions due to the normal state pseudogap (PG) and the
superconducting (SC) gap. They show that these phenomena involve different
energy scales, exhibit distinct doping dependencies and thus are likely of
different origin. In particular, the SW redistribution in the PG state
closely resembles the one of a conventional charge- or spin density wave
(CDW\ or SDW) system.
	
\end{abstract}

\maketitle

It is now widely recognized that the anomalous normal state electronic
properties of the cuprate high\ temperature superconductors (SC) may hold
the key for understanding the SC pairing mechanism. The most challenging
feature is the so-called pseudogap (PG) phenomenon which prevails in samples
that are underdoped with respect to the dome-shaped dependence of the SC
critical temperature, T$_{c}$, on hole doping, {\it p} \cite%
{Timusk99,Tallon01}. Here it gives rise to a gap-like depletion of the
low-energy charge and spin excitations starting at temperatures well above T$%
_{c}$. The PG has been identified by several experimental techniques like
specific heat \cite{Loram93}, angle-resolved photo-emission (ARPES) \cite%
{Loeser96}, c-axis tunneling \cite{Krasnov00}, and also by infrared (IR)
spectroscopy \cite{Homes93,Basov05,Bernhard99}. While these have established
some important aspects, like its strong k-space anisotropy \cite{Loeser96}
or the rapid increase of its magnitude towards the underdoped side \cite%
{Loeser96,Bernhard99}, the origin of the PG and, in particular, the question
of whether it is intrinsic or extrinsic with respect to SC remains heavily
debated \cite{Tallon01,Loram93,Kugler01,Miyakawa01}. Prominent intrinsic
theories are the phase fluctuation model where the PG corresponds to a SC
state lacking macroscopic phase coherence \cite{Emery95}, precursor pairing
models where the pair formation occurs at much higher temperature ($T)$ than
the condensation \cite{Kosztin00}, and the resonating valence bond (RVB)
theory \cite{Anderson04}. The extrinsic models include numerous conventional
and exotic spin- or charge density wave states that are either passive or
even competitive with respect to SC \cite{Chubukov98}. One of the major
obstacles in identifying the relevant theory is the lack of a clear
spectroscopic distinction between the PG and the SC gap.

In this letter we present broad-band infrared (70-4000 cm$^{-1}$)
ellipsometry measurements of the c-axis conductivity for a series of
underdoped RBa$_{2}$Cu$_{3}$O$_{7-\delta }$ (R=Y, Nd, La) single crystals
which shed new light on this issue. Our data detail the spectral weight (SW)
redistribution due to the formation of the normal state PG\ and the SC gap.
They highlight significant differences in the related energy scales and
their doping dependencies which provide evidence in favor of the extrinsic
PG models.

High quality RBa$_{2}$Cu$_{3}$O$_{7-\delta }$ (R=Y, Nd, La) single crystals
were grown with a flux method in Y-stabilized zirconium crucibles under
reduced oxygen atmosphere to avoid the substitution of R ions on the Ba site %
\cite{Schlachter00}. The oxygen content and the consequent hole doping of
the CuO$_{2}$ planes, {\it p}, were adjusted by annealing at appropriate $T$
in flowing O$_{2}$ gas and subsequent rapid quenching. The quoted T$%
_{c}(\Delta $T$_{c})$ values correspond to the midpoint (10 to 90 \% width)
of the diamagnetic SC transition as measured by SQUID magnetometry. The
doping levels were deduced from the empirical relationship \cite{Tallon01} 
{\it p}=0.16$\pm \sqrt{\frac{1-T_{c}/T_{c,\max }}{82.6}}$\ with $T_{c,max}$
= 92.5, 96 and 98 K for Y, Nd and La, respectively \cite{GVM96}. In
agreement with previous reports, we find that the oxygen content needed to
obtain a certain value of {\it p }increases with the size of the R ion \cite%
{GVM96}. The annealing temperatures (in pure O$_{2}$ gas), the resulting
oxygen deficiencies, $\delta $, and the T$_{c}$ values at $p\approx $0.12
are 570 $%
{{}^\circ}%
$C, $\delta \approx $0.2, T$_{c}$=82(1) K for Y-123, 440 $%
{{}^\circ}%
$C, $\delta \approx $0.1, T$_{c}$=85(2) K for Nd-123, and 310 $%
{{}^\circ}%
$C, $\delta \approx 0,$ T$_{c}$=87(2)K for La-123.

The ellipsometry experiments were performed with a home-built ellipsometer
attached to a Bruker Fast-Fourier spectrometer at the infrared (IR) beamline
at the ANKA synchrotron at FZ Karlsruhe, D for the range of 70-700 cm$^{-1}$
and with a corresponding laboratory-based setup at 400-4000 cm$^{-1}$ \cite%
{Bernhard04}. Ellipsometry directly measures the complex dielectric
function, $\widetilde{\varepsilon }=\varepsilon _{1}+i\varepsilon _{2}$, and
the related optical conductivity $\widetilde{\sigma }\left( \omega \right)
=i\cdot \omega \ (1-\widetilde{\varepsilon }\left( \omega \right) ),$
without a need for a Kramers-Kronig analysis \cite{Azzam77}. It is
furthermore a self-normalizing technique that enables a very accurate and
reproducible measurement of $\widetilde{\varepsilon }$ and, in particular,
of its {\it T}-dependent changes. We note that thanks to the large probe
depth of the infrared radiation ($\sim $1 $\mu $m) which ensures the bulk
nature of the observed phenomena, the high energy resolution, and the
accuracy in the determination of the optical constants which enables the
application of powerful sum rules, this technique provides important
complementary information with respect to other techniques like ARPES or
tunneling.

Figure 1 summarizes our data of the c-axis optical conductivity, $\sigma
_{1c}\left( \omega \right) ,$ for the underdoped (p$\approx $0.12) NdBa$_{2}$%
Cu$_{3}$O$_{6.9}$ crystal. Representative spectra in the PG and the SC
states are shown in Fig.1a and 1b, respectively. The spectral shape of the 
{\it T}-dependent changes is detailed in Fig.1c in terms of the conductivity
difference spectra, $\Delta \sigma _{1c}=\sigma _{1c}\left( T\right) -\sigma
_{1c}\left( 300\ K\right) $. The $T$-dependence of the integrated spectral
weight, SW$_{\alpha }^{\beta }$=$\stackrel{\beta }{%
\mathrel{\mathop{\int }\limits_{\alpha }}%
}\sigma _{1c}\left( \omega \right) d\omega $, is given in Figs. 1d-1f for
representative integration limits.

First, we focus on the spectral changes due to the normal state PG which
gives rise to a gap-like suppression of the low frequency part of $\sigma
_{1c}(\omega )$ below an onset temperature, T$^{\ast }>$T$_{c}$. From the 
{\it T}-dependence of SW$_{0^{+}}^{800}$ in\ Fig. 1d we estimate $T^{\ast
}\approx 220\ K$. Figure 1a shows the corresponding crossing point in $%
\sigma _{1c}\left( \omega ,T\right) $ at $\omega _{PG}^{\ast }\approx 700$ cm%
$^{-1}$ which separates the low frequency region, where $\sigma _{1c}$
undergoes a gap-like suppression, from the high frequency one, where $\sigma
_{1c}$ exhibits a corresponding increase. In classical spin- or charge
density wave (SDW or CDW) systems this crossing is determined by the
magnitude of the gap, i.e. $\hbar \omega ^{\ast }\approx $2$\Delta $ \cite%
{Gruener94}. In analogy, we use the same relationship to estimate the
magnitude of the PG, i.e $\hbar \omega _{PG}^{\ast }\approx $2$\Delta ^{PG}$%
. Even though the origin of the PG\ remains unknown, we still believe that
this estimate yields a reasonable value for $\Delta ^{PG}$ and especially
for its variation with doping. Apart from some controversy regarding the
determination of the magnitude of $\omega _{PG}^{\ast }$, which requires a
very precise measurement of $\Delta \sigma _{1c}(T)$, our data agree well
with previous reports of FIR\ reflectivity \cite{Timusk99,Homes93,Basov05}
and ellipsometry \cite{Bernhard99,Pimenov05}. In addition, our new
broad--band ellipsometry data answer the long standing question of where the
SW is accumulated that is removed from the FIR range (at $\omega <\omega
_{PG}^{\ast })$. Note that the SW, if integrated to a sufficiently high
frequency, should be conserved and thus {\it T} independent (the so-called
SW sum rule). Our data establish that the gap like suppression of $\sigma
_{1c}\left( \omega <\omega _{PG}^{\ast }\right) ,$ is balanced by the
corresponding increase of $\sigma _{1c}\left( \omega >\omega _{PG}^{\ast
}\right) $ which yields a broad MIR band that starts right at the gap edge
and extends to about 3500 cm$^{-1}$. The amount of transferred normal state
SW is $\sim $13000 $\Omega ^{-1}$cm$^{-2}$. The SW balance is evident from
the smooth evolution of SW$_{0^{+}}^{4000}$ which, unlike SW$_{0^{+}}^{800},$
does not exhibit a sizeable decrease below $T^{\ast }$. The remaining
decrease of SW$_{0^{+}}^{4000}$ (mostly above 200 K) is actually unrelated
to the PG phenomenon. It occurs for all doping levels both for the c-axis
(data for overdoped samples are not shown here)\ and for the in-plane
response \cite{Ortolani05}. Figure 1c also shows that the related $\Delta
\sigma _{1c}\left( \omega \right) $ spectra are rather different, i.e. $%
\Delta \sigma _{1c}$($\omega $) is always positive at {\it T}%
\mbox{$>$}%
{\it T}$^{\ast }$ while at {\it T}$_{c}$%
\mbox{$<$}%
{\it T%
\mbox{$<$}%
T}$^{\ast }$ it exhibits a sign change at $\omega _{PG}^{\ast }$. Notably,
the spectral shape of the PG-related SW redistribution closely resembles the
one of a conventional charge- or spin density wave (CDW\ or SDW) system
which is described in Ref. \cite{Gruener94}.

Next we discuss what impact the SC transition has on the PG-related SW
redistribution, in particular, on the broad MIR band. We focus here on the
data at $\omega $%
\mbox{$>$}%
1600 cm$^{-1}$ which are not affected by the much narrower and SC-induced
band near 1000 cm$^{-1}$ which is unrelated to the PG phenomenon as
discussed below. In the first place, Fig. 1b shows that $\sigma _{1c}$($%
\omega >1600$ cm$^{-1})$ undergoes hardly any noticeable changes below $%
T_{c} $. Any SC induced modification of the broad MIR band therefore must be
fairly small. Nevertheless, a weak but yet significant impact of SC on the
MIR band is apparent from the $T$ dependence of SW$_{1600}^{4000}$ in Fig.
1f. Its rapid increase below $T^{\ast }$ is suddenly interrupted near $T_{c}$
followed by a saturation or even a minor decrease below {\it T}$_{c}$. We
have confirmed this trend for several underdoped Y-123 crystals (data are
not shown here). Notably, a similar behavior was previously observed in the
A-15 compounds where the onset of SC arrests the anomalous softening of a
phonon mode that is associated with a CDW-like\ instability \cite{Testardi75}%
. Just like this observation had inspired the classical theoretical work on
competing CDW (SDW) and SC order parameters \cite{Balseiro79}, we hope that
our new results will motivate further theoretical studies concerning the SW
redistributions in the context of the various proposed PG models. We also
note that our data seem to provide a challenge to the phase fluctuation
models. Naively one would expect here that a significant part of the SW of
the broad MIR band is transferred back to low energies and joins the SC
condensate once the phase fluctuations diminish below $T_{c}$. Not to be
misunderstood, we do not argue that our data provide evidence against the
existence of precursor SC fluctuations, which are meanwhile well
established. However, our data suggest that these SC fluctuations are not
causing the normal state PG phenomenon. Instead they may well be a
consequence of the PG-like suppression of the low energy SW and thus of the
SC condensate density. The minor decrease of SW$_{0^{+}}^{4000}$ below about
120 K in Fig. 1e may well be a signature of these precursor SC fluctuations.

Finally, we note that our data confirm previous reports \cite{Basov99} that
the SW transfer in the SC state involves an anomalously high energy scale.
This is evident from Fig. 1e where SW$_{0}^{4000},$ including the SW of the
SC delta function at the origin, SW$^{\delta }$ (solid blue circles, as
deduced by fitting a Drude-Lorentz model to the complex dielectric
function), apparently exhibits an anomalous increase below\ T$_{c}$. The
comparison with the estimated trend based on a power law fit to the normal
state data, as shown by the red thin line in Fig. 1e, suggests that about 20
\% of SW$^{\delta }$ originates from the high frequency range of $\omega >$%
4000 cm$^{-1}$. Notably, our data suggest that this anomalous contribution
to SW$^{\delta }$ does not arise from a partial reversal of the PG-related
SW transfer which is confined to the region below 4000 cm$^{-1}$ as shown
above.

In Fig. 2 we compare the doping dependencies of the spectral gaps in the PG
and SC\ states. In analogy to the PG case, we estimate the magnitude of the
SC gap, $\Delta ^{SC}$, as 2$\Delta ^{SC}\approx \hbar \omega _{SC}^{\ast }$%
, where $\omega _{SC}^{\ast }$ is the frequency of the crossing point as
shown in Figs. 1b and 2c. A justification for this approach is given in the
next paragraph. Figure 2a displays the resulting phase diagram of $\Delta
^{PG}$ and $\Delta ^{SC}$ as obtained for a series of strongly to very
weakly underdoped Y-123 single crystals. The evaluation of $\omega
_{PG}^{\ast }$ and $\omega _{SC}^{\ast }$ is detailed in Figs. 2b and 2c,
respectively. In the first place, Fig. 2a shows that $\Delta ^{PG}$ exhibits
a considerably stronger doping dependence than $\Delta ^{SC}.$ Notably,
there is a crossing point around p$\approx $0.12 which separates regions of $%
\Delta ^{PG}$ 
\mbox{$>$}%
$\Delta ^{SC}$ ($\Delta ^{PG}$ 
\mbox{$<$}%
$\Delta ^{SC})$ at low (high) doping. We note that the finding of $\Delta
^{PG}$ 
\mbox{$>$}%
$\Delta ^{SC}$ for {\it p}%
\mbox{$<$}%
0.12 has important implications, in particular, it contrasts with the
expectation that the PG and the SC\ gap add to form the measured
spectroscopic gap, i.e. that $\Delta =\sqrt{\Delta _{PG}^{2}+\Delta _{SC}^{2}%
}$. The observation of $\Delta ^{PG}>\Delta ^{SC}$ may be accounted for in
terms of a spatial separation of the two gaps. However, our optical data can
not tell whether this separation takes place in real or in k-space.
Nevertheless, the latter possibility is indeed suggested by recent ARPES
data \cite{Tanaka06}. Furthermore, while we cannot follow the evolution of $%
\Delta ^{PG}$ beyond optimum doping where $T^{\ast }\leq T_{c}$, the linear
extrapolation of our data suggests that the PG vanishes much earlier than $%
\Delta ^{SC}$ at a critical doping of p$_{crit}\approx 0.19-0.2$ which
agrees well with previous reports of a sudden change in the electronic
properties \cite{Tallon01,Pimenov05}. Accordingly, our data once more favor
the extrinsic PG models, in particular, they suggest that $\Delta ^{PG}$ and 
$\Delta ^{SC}$ do not merge on the overdoped side as predicted by the
intrinsic models.

We return now to the discussion of the SC-induced mode near 1000 cm$^{-1}$
mode and the relationship between $\hbar \omega _{SC}^{\ast }$ and 2$\Delta
^{SC}$. Figure 3 shows data for a series of similarly underdoped (p$\approx $%
0.12) R-123 with R=Y, Nd, and La crystals which establish that the intensity
of the 1000 cm$^{-1}$ strongly depends on the size of the ionic radius of
the R ion and the related strength of the electronic bilayer coupling. The
expected correlation between the R ionic size which increases from 1.019
\AA\ for Y$^{3+},$ to 1.109 \AA\ for Nd$^{3+},$ and 1.16 \AA\ for La$^{3+}$\ %
\cite{GVM96} and the bilayer coupling is confirmed by recent ARPES\
measurements on the same samples (private communication with S. Borisenko)
and also by the sizeable red shift of the so-called transverse Josephson
plasma resonance (t-JPR) mode \cite{vdM97} near 500 cm$^{-1}$. Here we only
mention two possible interpretations of the 1000 cm$^{-1}$ mode. Firstly, it
may correspond to an interband (bonding-antibonding) pair-breaking peak
whose coherence factor decreases with decreasing bilayer coupling \cite%
{Munzar06}. This appealing interpretation is still speculative because final
state interactions have not been taken into account in the calculations of
Ref. \cite{Munzar06}. Alternatively, the mode could be due to final states
involving two Bogulyubov quasiparticles and a bosonic mode (e.g. the neutron
resonance), an interpretation similar to that of the well known maximum
around 1000 cm$^{-1}$ in the SC state in-plane conductivity \cite{Basov05}.
Irrespective of the outcome of this ongoing work, we note that according to
both interpretations the maximum is located a few tens meV above the SC\
gap, suggesting that the difference between $\Delta ^{SC}$ and the true
single particle gap is small. This conclusion is furthermore supported by
the data of the La-123 crystal which are shown in Fig.4. Here the 1000 cm$%
^{-1}$ mode is absent and 2$\Delta ^{SC}$ can be estimated as $\hbar \omega
_{SC}^{\ast }$/2 \cite{Hirschfeld96}, where $\omega _{SC}^{\ast }$ is the
frequency of the shoulder feature as marked by the cyan arrow. Notably, the
obtained value of $2\Delta ^{SC}\approx 800$ cm$^{-1}$ compares rather well
with the ones of $2\Delta ^{SC}\approx 780$ cm$^{-1}$ in Y-123 and $2\Delta
^{SC}\approx 860$ cm$^{-1}$ in Nd-123. Another interesting trend occurs for
the PG magnitude of $2\Delta ^{PG}\approx 1200$ cm$^{-1}$ in La-123 which is
almost doubled as compared to the one in Y-123 and Nd-123. At present we do
not know whether this trend reflects a strong dependence of the PG
correlations on the electronic bilayer coupling or rather a sensitivity to
structural disorder, e.g. in the Ba-O layer, which is likely enhanced in
La-123. This issue deserves further investigation and may well open up new
possibilities for distinguishing between the PG phenomenon SC gap with other
spectroscopic techniques.

We acknowledge financial support by the Schweizer Nationalfonds (SNF) by
grant 200021-111690/1 and the Deutsche Forschungsgemeinschaft (DFG) by grant
BE2684/1-1 in FOR538. D.M. and J.C were supported by the Ministry of
Education of CR (MSM0021622410). We are indepted to Y.L. Mathis for
technical support at ANKA.

Figure 1: Infrared c-axis conductivity, $\sigma _{1c}$($\omega $), of
underdoped NdBa$_{2}$Cu$_{3}$O$_{6.9}$. Shown are representative spectra for
(a) the PG state, (b) the SC state, and (c) the {\it T}-difference, $\Delta
\sigma _{1c}=\sigma _{1c}\left( T\right) -\sigma _{1c}\left( 300\ K\right) $%
. Green (cyan) arrows mark the crossing points at $\omega _{PG}^{\ast }$ ($%
\omega _{SC}^{\ast })$. Also shown is the $T$-dependence of the integrated
spectral weight, SW$_{\alpha }^{\beta }$= $\stackrel{\beta }{%
\mathrel{\mathop{\int }\limits_{\alpha }}%
}\sigma _{1c}\left( \omega \right) d\omega ,$ for (d) $\alpha $=0$^{+}$ and $%
\beta $=800 cm$^{-1}$ (regular part), (e) $\alpha $=0 and $\beta $=4000 cm$%
^{-1}$ (black squares show the regular part while blue circles include the
contribution of the SC delta function, SW$^{\delta }),$ and (f) $\alpha $%
=1600 to $\beta $=4000 cm$^{-1}$. The red line in (e) shows a fit to the
normal state data with the function SW$\left( T\right) $=SW$_{0}\left(
1-\left( \frac{T}{\Theta }\right) ^{\beta }\right) .$

Figure 2: (a) Phase diagram of Y-123 showing the doping dependencies of the
critical temperature, $T_{c}$ (black stars), the PG\ magnitude,$\Delta ^{PG}$
(green circles), and the SC\ gap, $\Delta ^{SC}$ (blue triangles). The
dashed green (cyan) line shows a linear extrapolation of $\Delta ^{PG}$ ($%
\Delta ^{SC}$) towards the overdoped side. (b) and (c) Conductivity
difference spectra, $\Delta \sigma _{1c}^{PG}=\sigma _{1c}\left( T\approx
T_{c}\right) -\sigma _{1c}\left( T\approx T^{\ast }\right) $ and $\Delta
\sigma _{1c}^{SC}=\sigma _{1c}\left( 10K\right) -\sigma _{1c}\left( T\approx
T_{c}\right) ,$ showing the changes due to the PG and the SC gap,
respectively. Some sharp features due to a $T$-dependent shift of phonon
modes have been removed for clarity.

Figure 3: (a)-(c) Infrared c-axis conductivity for R-123 (R=Y, Nd and La)
with {\it p}$\approx $0.12. Black dotted (solid) arrows mark the transverse
Josephson plasma mode (mode near 1000 cm$^{-1})$. (d) Corresponding
difference spectra, $\Delta \sigma _{1c}=\sigma _{1c}\left( 10K\right)
-\sigma _{1c}\left( 90K\right) .$

Figure 4: Infrared c-axis conductivity, $\sigma _{1c}$($\omega $), of
underdoped LaBa$_{2}$Cu$_{3}$O$_{7}$ with T$_{c}$=87(2) K. Green and cyan
arrows mark the onset of the PG and the SC gap, respectively.

\end{document}